\documentclass{ws-procs9x6}

\newcommand{\zr}[1]{\mbox{\hspace*{#1em}}}

\newcommand{\ZZ}{\mbox{\sf Z\zr{-0.45}Z}}

\newcommand{\Ref}[1]{Ref.~\refcite{#1}}

\newcommand{\Hs}{\hspace*{1em}}

\newcommand{\Eq}[1]{Eq.~(\ref{#1})}
\newcommand{\Fig}[1]{Fig.~\ref{#1}}
\newcommand{\Table}[1]{Table~\ref{#1}}

\begin{document}

\title{Anisotropic lattice QCD studies of penta-quark anti-decuplet
\footnote{\uppercase{L}attice  \uppercase{QCD}  numerical  calculation
has    been   done    with    \uppercase{NEC}   \uppercase{SX}-5    at
\uppercase{O}saka \uppercase{U}niversity.}  }

\author{N.~Ishii$^1$, T.~Doi$^2$, H.~Iida$^1$,
M.~Oka$^1$, F.~Okiharu$^3$,  H.~Suganuma$^1$}

\address{
$^1$ Dept. of Phys., Tokyo Institute of Technology, Meguro, Tokyo 152-8551, Japan\\
$^2$ RIKEN BNL Research Center, BNL,
Upton, New York 11973, USA\\
$^3$ Faculty of Science and Tech., Nihon Univ.,
Chiyoda, Tokyo 101-8308, Japan}





\maketitle

\vspace*{-2ex}
\abstracts{  Anti-decuplet  penta-quark  baryon  is studied  with  the
quenched  anisotropic  lattice QCD  for  accurate  measurement of  the
correlator. Both  the positive and negative parity  states are studied
using a non-NK type interpolating  field with $I=0$ and $J=1/2$. After
the chiral extrapolation, the lowest positive parity state is found at
$m_{\Theta}\simeq  2.25$ GeV, which  is too  massive to  be identified
with the experimentally observed $\Theta^+(1540)$. The lowest negative
parity state is found at  $m_{\Theta}\simeq 1.75$ GeV, which is rather
close to the empirical value.
To confirm  that this state  is a compact  5Q resonance, a  new method
with ``{\em  hybrid boundary condition (HBC)}'' is  proposed.  The HBC
analysis shows that the observed  state in the negative parity channel
is an NK scattering state.  }
\vspace*{-8ex}

\section{Introduction}
LEPS   group   at  SPring-8   has   discovered   a  narrow   resonance
$\Theta^+(1540)$,  which is  centered at  $1.54 \pm  0.01$ GeV  with a
width smaller than $25$ MeV.\cite{nakano}
This resonance is confirmed to have baryon number $B=1$, charge $Q=+1$
and  strangeness $S=+1$  implying that  it is  a baryon  containing at
least   one   $\bar{s}$.   Hence,   its   simplest  configuration   is
$uudd\bar{s}$, i.e., a manifestly exotic penta-quark (5Q) state.
The  experimental   discovery  of   $\Theta^+$  was  motivated   by  a
theoretical prediction.\cite{diakonov}

Tremendous theoretical  efforts have been and are  still being devoted
to  the  investigation of  $\Theta^+$,\cite{oka,zhu}  among which  its
parity  is one  of  the most  important  topics.  Experimentally,  the
parity       determination      of      $\Theta^+$       is      quite
challenging,\cite{hicks,thomas}  while  opinions  are divided  in  the
theoretical side.\cite{oka}

There   are  several   quenched  lattice   QCD  studies   of   the  5Q
state,\cite{fodor,sasaki,chiu,lee} However,  the results have  not yet
reached a consensus.
One  group \cite{chiu} claims  the existence  of a  low-lying positive
parity 5Q resonance.   Negative parity 5Q resonance is  claimed by two
groups,\cite{fodor,sasaki}  among  which  \Ref{sasaki} has  omitted  a
quark-exchange  diagram  between  diquark  pairs assuming  the  highly
correlated diquark picture.
Note  that  these  three  groups employed  non-NK  type  interpolating
fields.
In contrast,  \Ref{lee} has employed the  NK-type interpolating field,
and  performed solid  analysis  concluding  that no  signal  for a  5Q
resonance state is observed.
There  is  another  type of  lattice  QCD  studies  of the  static  5Q
potential  \cite{okiharu} aiming at  providing physical  insights into
the structure of 5Q baryons.

In this paper, we study the  5Q baryon $\Theta^+$ for both parities by
using  high-precision  data generated  with  the quenched  anisotropic
lattice  QCD.    We  employ  the  standard  Wilson   gauge  action  at
$\beta=5.75$  on the  $12^3\times  96$ lattice  with the  renormalized
anisotropy  $a_s/a_t  =  4$.   The  anisotropic lattice  method  is  a
powerful  technique, which  can  provide us  with high-precision  data
quite efficiently.\cite{klassen,matsufuru,nemoto,ishii}
The  lattice spacing  is determined  from the  static  quark potential
adopting the Sommer parameter $r_0^{-1}= 395$ MeV leading to $a_s^{-1}
= 1.100(6)$ GeV ($a_s  \simeq 0.18$ fm).\cite{matsufuru}
The lattice size $12^3\times  96$ amounts to $(2.15\mbox{fm})^3 \times
4.30\mbox{fm}$ in the physical unit.
The   $O(a)$-improved  Wilson  quark   (clover)  action   is  employed
\cite{matsufuru}   with   four  values   of   hopping  parameters   as
$\kappa=0.1210(0.0010)0.1240$, which  roughly covers $m_s  \le m_q \le
2m_s$  corresponding  to $m_{\pi}/m_{\rho}  =  0.81,  0.77, 0.72$  and
$0.65$.
By   keeping  $\kappa_s=0.1240$   fixed   for  s   quark,  we   change
$\kappa=0.1210-0.1240$ for u and  d quarks for chiral extrapolation.
Unless otherwise indicated, we use
\begin{equation}
  (\kappa_s,\kappa)=(0.1240,0.1220),
\label{typical.set}
\end{equation}
as a typical set of hopping parameters.
Anti-periodic boundary  condition (BC) is imposed on  the quark fields
along the temporal direction.
To enhance the  low-lying spectra, we adopt a  smeared source with the
gaussian size $\rho\simeq 0.4$ fm.
We  use   504  gauge   configurations  to  construct   correlators  of
$\Theta^+$.
For detail, see \Ref{ishii-penta}.

In the former part of this  paper, we present the standard analysis of
5Q correlators in  both the positive and the  negative parity channels
adopting  the  standard  periodic  boundary  condition  along  spatial
directions.
Latter half of this paper is devoted to a further investigation of the
negative  parity state.  Proposing  a new  general method  with ``{\em
hybrid boundary condition (HBC)}'', we attempt to determine whether it
is a compact resonance state or a NK scattering state.
\section{Parity projection}
We consider a non-NK type interpolating field for $\Theta^+$ as
\begin{equation}
  O
  \equiv
  \epsilon_{abc}
  \epsilon_{ade}
  \epsilon_{bfg}
  \left(u_d^T C\gamma_5 d_e\right)
  \left(u_f^T C d_g\right)
  \left(C\bar{s}_c^T\right),
  \label{sasaki-op}
\end{equation}
where  $a-g$  denote  color  indices, and  $C\equiv  \gamma_4\gamma_2$
denotes the charge conjugation matrix.
The quantum number of $O$ is spin $J=1/2$ and isospin $I=0$.
Under the spatial  reflection of the quark fields,  i.e., $q(t,\vec x)
\to \gamma_4  q(t,-\vec x)$, $O$  transforms exactly in the  same way,
i.e., $O(t,\vec x)  \to +\gamma_4 O(t,-\vec x)$, which  means that the
intrinsic parity of $O$ is positive.
Although  its intrinsic  parity is  positive, it  couples  to negative
parity states as well.\cite{montvay}

We consider  the asymptotic  behavior of the  correlator in the  5Q CM
frame as
\begin{equation}
  G_{\alpha\beta}(t)
  \equiv
  \frac1{V}
  \sum_{\vec x}
  \left\langle
  O_{\alpha}(t,\vec x)
  \bar{O}_{\beta}(0,\vec 0)
  \right\rangle,
\end{equation}
where $V$ denotes  the spatial volume.  In the region of  $0 \ll t \ll
N_t$  ($N_t$ : temporal  lattice size),  the correlator  is decomposed
into two parts as
\begin{eqnarray}
  G(t)
  &\equiv&
  P_{+}
  \left( C_+ e^{-m_+ t} - C_- e^{-m_-(N_t - t)} \right)
  \label{spectral.rep}
  \\\nonumber
  &+&
  P_{-}
  \left( C_- e^{-m_- t} - C_+ e^{-m_+(N_t - t)} \right),
\end{eqnarray}
where  $m_{\pm}$  refer to  the  energies  of  lowest-lying states  in
positive and  negative parity channels,  respectively. $P_{\pm} \equiv
(1\pm \gamma_4)/2$ serve as projection matrices onto the ``upper'' and
``lower''  Dirac  subspaces,   respectively,  in  the  standard  Dirac
representation.   \Eq{spectral.rep} suggests  that, in  the  region of
$0\ll  t  \ll  N_t/2$,   the  backwardly  propagating  states  can  be
neglected.   Hence,  ``upper''  Dirac  subspace is  dominated  by  the
lowest-lying positive  parity state, whereas  ``lower'' Dirac subspace
is dominated  by the lowest-lying  negative parity state.   We utilize
this property in parity projection.
%
\section{Numerical Result with standard BC}
\begin{figure}
\vspace*{-4ex}
\begin{center}
\begin{tabular}{cc}
\psfig{file=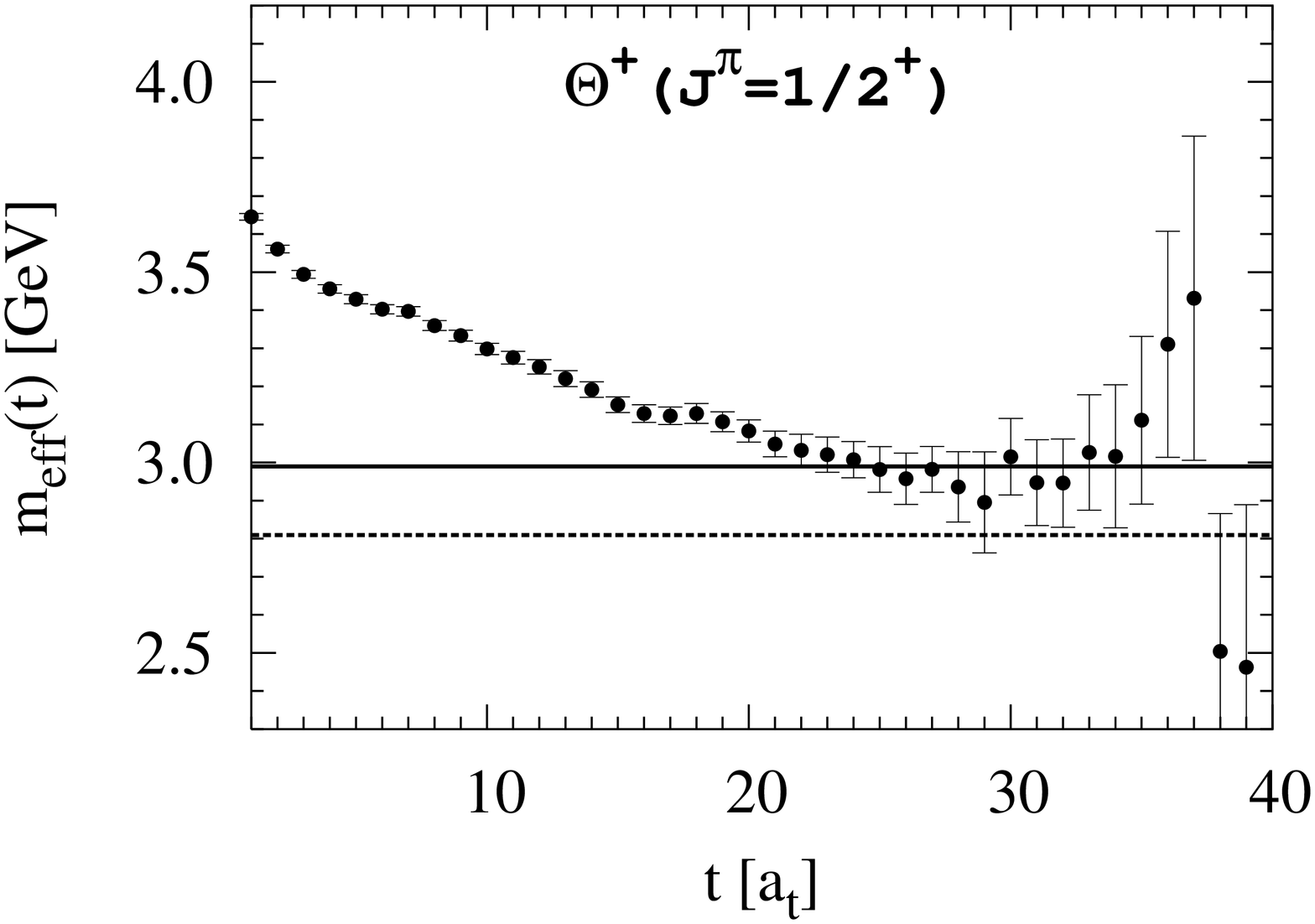,width=0.48\textwidth,angle=-90}
&
\psfig{file=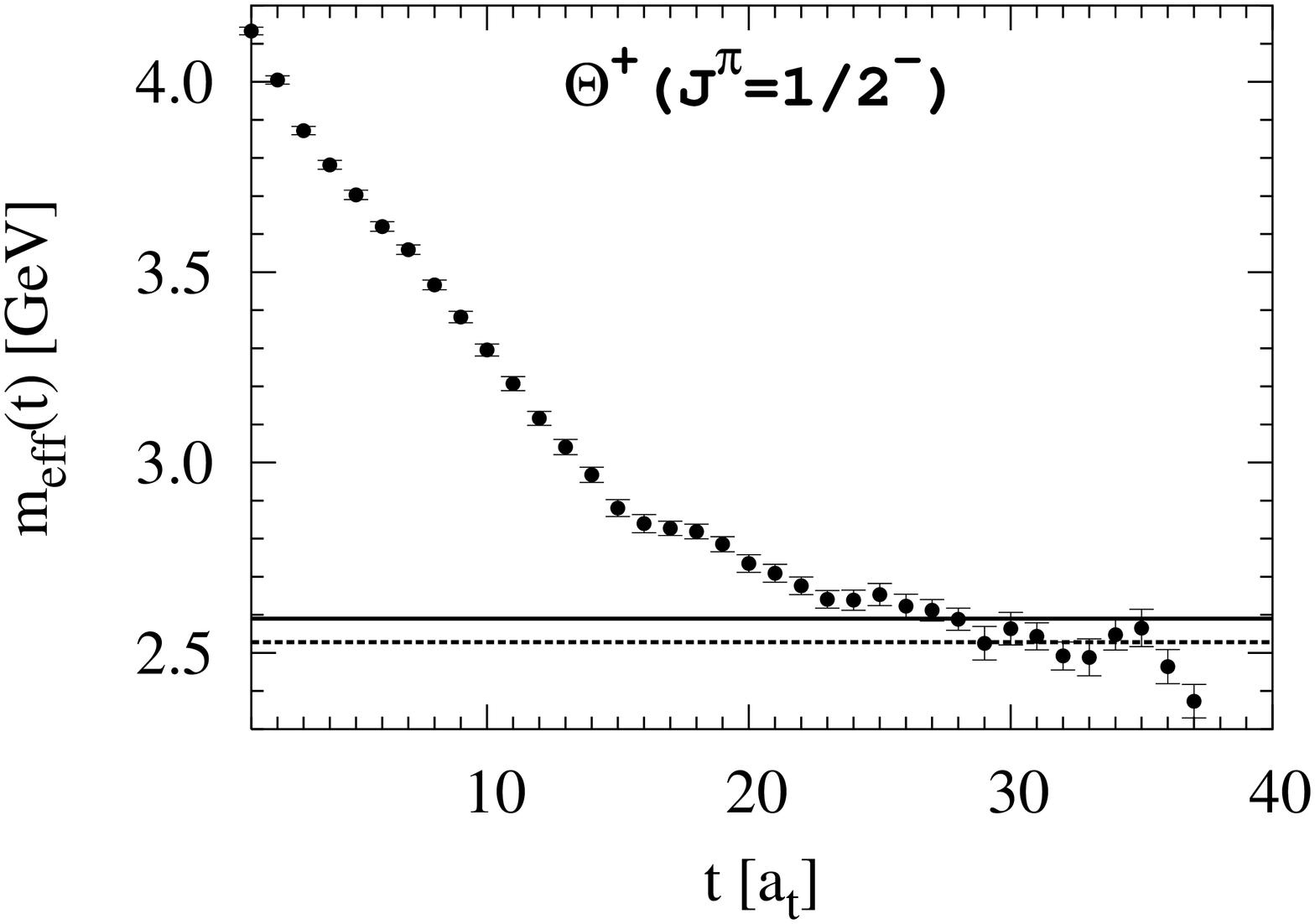,width=0.48\textwidth,angle=-90}
\end{tabular}
\end{center}
\vspace*{-2ex}
\caption{The  effective mass  plots  of positive  and negative  parity
$\Theta^+$  adopting  \Eq{typical.set}.  The  solid  lines denote  the
result of the single-exponential fit performed in the region, $25\le t
\le  35$. The  dotted lines  denote the  p-wave (s-wave)  NK threshold
energy for positive (negative)  parity channels on the spatial lattice
size $L \simeq 2.15$ fm.}
\label{fig.effmass}
\vspace*{-2ex}
\end{figure}

In \Fig{fig.effmass},  we show the  effective mass plots for  both the
parity channels adopting \Eq{typical.set}.
The effective mass is defined as
\begin{equation}
  m_{\rm eff}(t)
  \equiv
  \log(g(t)/g(t+1)),
\end{equation}
where  $g(t)$  denotes  correlator  in Dirac  ``upper''  or  ``lower''
subspaces.    Formally,  $g(t)$  can   be  expressed   as  a   sum  of
exponentials.   In  the  asymptotic  region  $0\ll  t  \ll  N_t/2$  in
Euclidean time,  contaminations of excited  states are expected  to be
reduced.  If  $g(t)$ is dominated by  single exponential corresponding
to  the lowest-lying  mass $m$,  then  the effective  mass behaves  as
constant in  this region,  i.e., $m_{\rm eff}(t)  \sim m$.   Owing to
this property, the effective mass  plot is often used to determine the
fit range.\cite{montvay}

For both the parity channels, we find plateaus in the region $25 \le t
\le  35$,  where  single-exponential   dominance  is  expected  to  be
achieved.   We   simply  neglect  the   data  for  $t  >   35$,  where
contributions from  the backward propagations are seen  to become less
negligible.  The  single-exponential fit  is performed in  the plateau
region.  The  results are  denoted by solid  lines.  The  dotted lines
indicate  the p-wave  (s-wave) NK  thresholds for  positive (negative)
parity channels  on the spatial  lattice size $L\simeq 2.15$  fm. Note
that  due to the  quantized spatial  momentum in  the finite  box, the
p-wave threshold  is raised as  $E_{\rm th} \simeq \sqrt{m_N^2  + \vec
p_{\rm min}^2} + \sqrt{m_K^2 + \vec p_{\rm min}^2}$ with $|\vec p_{\rm
min}| = 2\pi/L$.

\begin{figure}
\centerline{\psfig{file=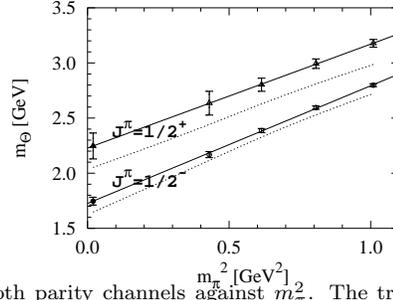,width=0.5\textwidth,angle=-90}}
\vspace*{-2.5ex}
\caption{$m_{\Theta}$    for     both    parity    channels    against
$m_{\pi}^2$. The  triangles correspond  to the positive  parity, while
the circles correspond to the negative parity. The open symbols denote
direct lattice  data, whereas  the closed ones  the results  after the
chiral  extrapolation.  The  dotted  lines indicate  the NK  threshold
energies for p-wave (upper) and s-wave (lower) cases.}
\label{fig.chiral}
\vspace*{-2ex}
\end{figure}

In \Fig{fig.chiral},  the masses  of positive (triangle)  and negative
(circle) parity  $\Theta^+$ are plotted against  $m_{\pi}^2$. The open
symbols denote  direct lattice  data.  We find  that the  data behaves
linearly in $m_{\pi}^2$. Such a linear behavior against $m_{\pi}^2$ is
also     observed      for     ordinary     non-PS      mesons     and
baryons.\cite{matsufuru,nemoto}   We  extrapolate  the   lattice  data
linearly to the  physical quark mass region.  The  results are denoted
by closed symbols.  For convenience, we show p-wave (upper) and s-wave
(lower) NK threshold with dotted lines.

In  the positive  parity channel,  the chiral  extrapolation  leads to
$m_{\Theta}=2.25$  GeV.   Since  it  is  too  massive,  it  cannot  be
identified with the experimentally observed $\Theta^+(1540)$.
In contrast, in the  negative parity channel, the chiral extrapolation
leads to $m_{\Theta}=1.75$ GeV, which is rather close to the empirical
value.
In order for this to be identified with $\Theta^+(1540)$, it should be
confirmed that the  observed state is not a NK  scattering state but a
compact 5Q resonance state. We  will pursue this direction in the next
section.
\section{Further investigation with hybrid BC}
In  the   positive  parity  channel,  NK  scattering   states  are  in
p-wave. Hence, in the 5Q CM frame,  the minimum momenta of N and K are
non-zero, i.e., $|\vec p_{\rm min}|  = 2\pi/L$ ($L$ : the spatial size
of the  lattice), through which  the NK threshold energy  acquires the
explicit volume dependence  as $ E_{\rm th} \simeq  \sqrt{m_N^2 + \vec
p_{\rm  min}^2} +  \sqrt{m_K^2 +  \vec p_{\rm  min}^2}$.  This  can be
utilized  to determine  whether  the  state of  concern  is a  compact
resonance or a NK scattering state.
On  the other  hand, in  the  negative parity  channel, NK  scattering
states are in s-wave. Hence, in the 5Q CM frame, N and K can have zero
spatial  momentum,  and the  NK  threshold  energy  does not  have  an
explicit volume  dependence, i.e., $E_{\rm th}\simeq m_N  + m_K$. This
is not convenient for our purpose.

We may  ask ourselves whether  there could be some  prescription which
can  provide  s-wave  NK  threshold  energy with  an  explicit  volume
dependence as the p-wave one.
%
\begin{table}[h]
\vspace*{-2ex}
\tbl{The {\em  hybrid boundary condition  (HBC)} imposed on  the quark
fields.  The  second  line   shows  the  standard  (periodic)  BC  for
comparison.}
{\footnotesize
\begin{tabular}{lccc}
\hline
	& u quark	& d quark 	& s quark \\
\hline
HBC	& anti-periodic	& anti-periodic	& periodic \\
standard BC	& periodic	& periodic	& periodic \\
\hline
\end{tabular}
\label{hbc}
}
\vspace*{-2ex}
\end{table}
This is  achieved by twisting  the spatial boundary condition  (BC) of
quark fields in  a flavor dependent manner as  follows.  We impose the
anti-periodic BC on u and d quark fields, while periodic BC on s quark
field.   We will  refer to  this boundary  condition as  ``{\em hybrid
boundary condition (HBC)}''. (See \Table{hbc}.)

Under HBC, hadrons are subject to their own spatial BC.  For instance,
since N$(uud, udd)$ and  K$(u\bar{s},d\bar{s})$ contain odd numbers of
u  and  d  quarks, they  are  subject  to  the anti-periodic  BC.   In
contrast, since $\Theta^+(uudd\bar{s})$ contains even numbers of u and
d quarks, it is subject to the periodic BC.  (See \Table{hbc-hadron}.)
\begin{table}[h]
\tbl{The consequence of the HBC on hadrons}
{\footnotesize
\begin{tabular}{lcccll}
\hline
	& quark content	& spatial BC	& minimum momentum & \\
\hline
N & $uud$,$udd$	& anti-periodic	& $(\pm\pi/L,\pm\pi/L,\pm\pi/L)$ & $|\vec p_{\rm min}| = \sqrt{3}\pi/L$ \\
K & $u\bar{s}$,$d\bar{s}$ & anti-periodic	& $(\pm\pi/L,\pm\pi/L,\pm\pi/L)$ & $|\vec p_{\rm min}| = \sqrt{3}\pi/L$ \\
$\Theta^+$ & $uudd\bar{s}$	& periodic & $(0,0,0)$ & $|\vec p_{\rm min}| = 0$\\
\hline
\end{tabular}
\label{hbc-hadron}
}
\vspace*{-2ex}
\end{table}
Recall  that, in  the box  of the  size $L$,  the spatial  momenta are
quantized as $p_i  = 2n_i\pi/L$ for periodic BC  and $(2n_i + 1)\pi/L$
for anti-periodic  BC with $n_i  \in \ZZ$.  Therefore,  $\Theta^+$ can
have zero spatial  momentum as $|\vec p_{\rm min}|=0$.  In contrast, N
and  K  have  non-zero   minimum  spatial  momenta  as  $|\vec  p_{\rm
min}|=\sqrt{3}\pi/L$.  Thus, under HBC,  s-wave NK threshold energy is
raised as
\begin{equation}
  E_{\rm th}
  \simeq
  \sqrt{m_N^2 + \vec p_{\rm min}^2}
  +
  \sqrt{m_K^2 + \vec p_{\rm min}^2},
  \Hs
  |\vec p_{\rm min}| = \sqrt{3}\pi/L,
\label{HBC.threshold}
\end{equation}
whereas a compact 5Q resonance state is expected to be unaffected.
  
In  order to  see that  HBC does  not affect  the  spatially localized
resonance  states, we  show  an example  of  an established  resonance
$\Sigma(uds)$ baryon.   We select $\Sigma$,  because it is  subject to
the periodic BC unlike N.
\begin{figure}
\begin{center}
\vspace*{-4ex}
\begin{tabular}{cc}
\psfig{file=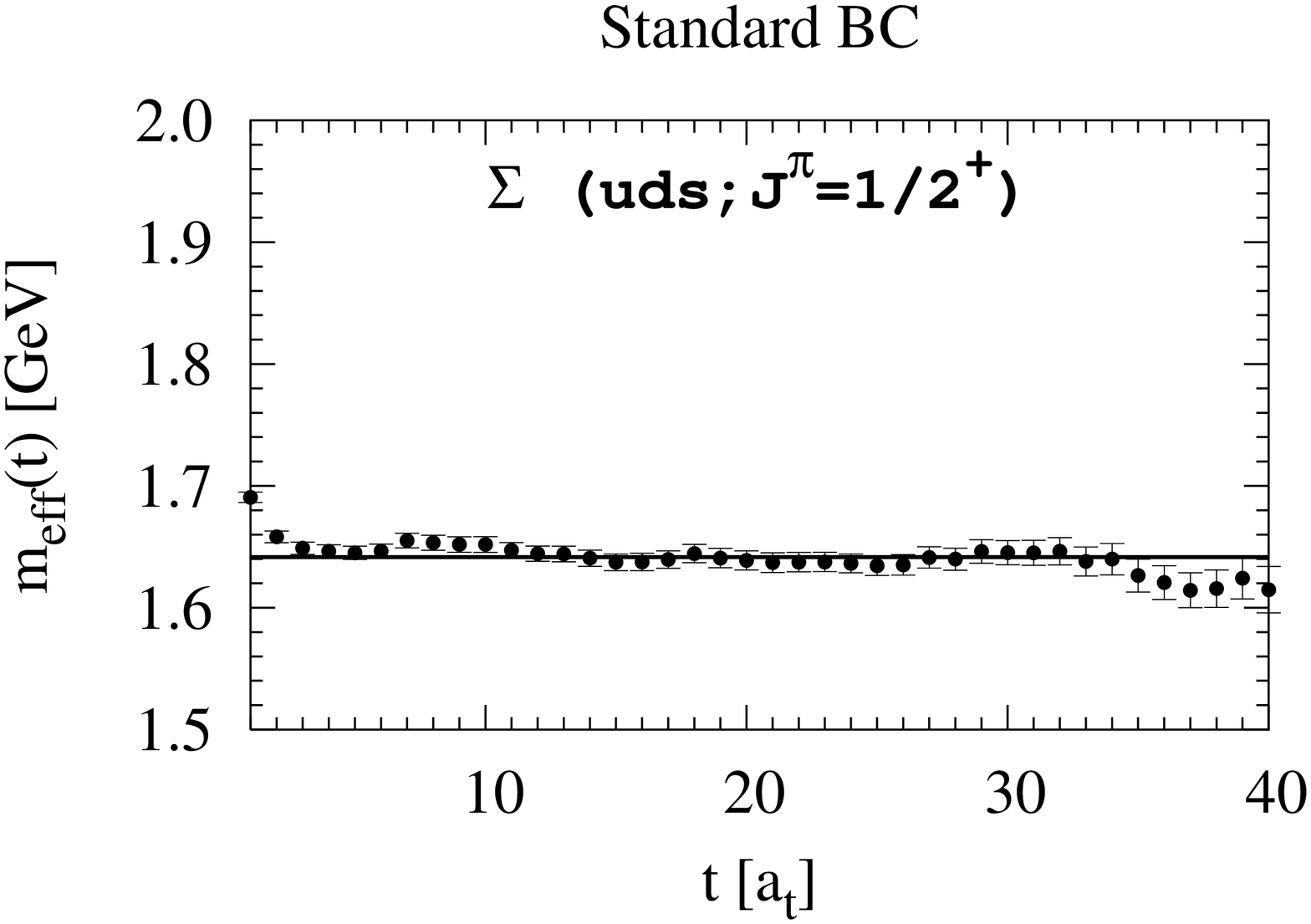,width=0.48\textwidth,angle=-90}
&
\psfig{file=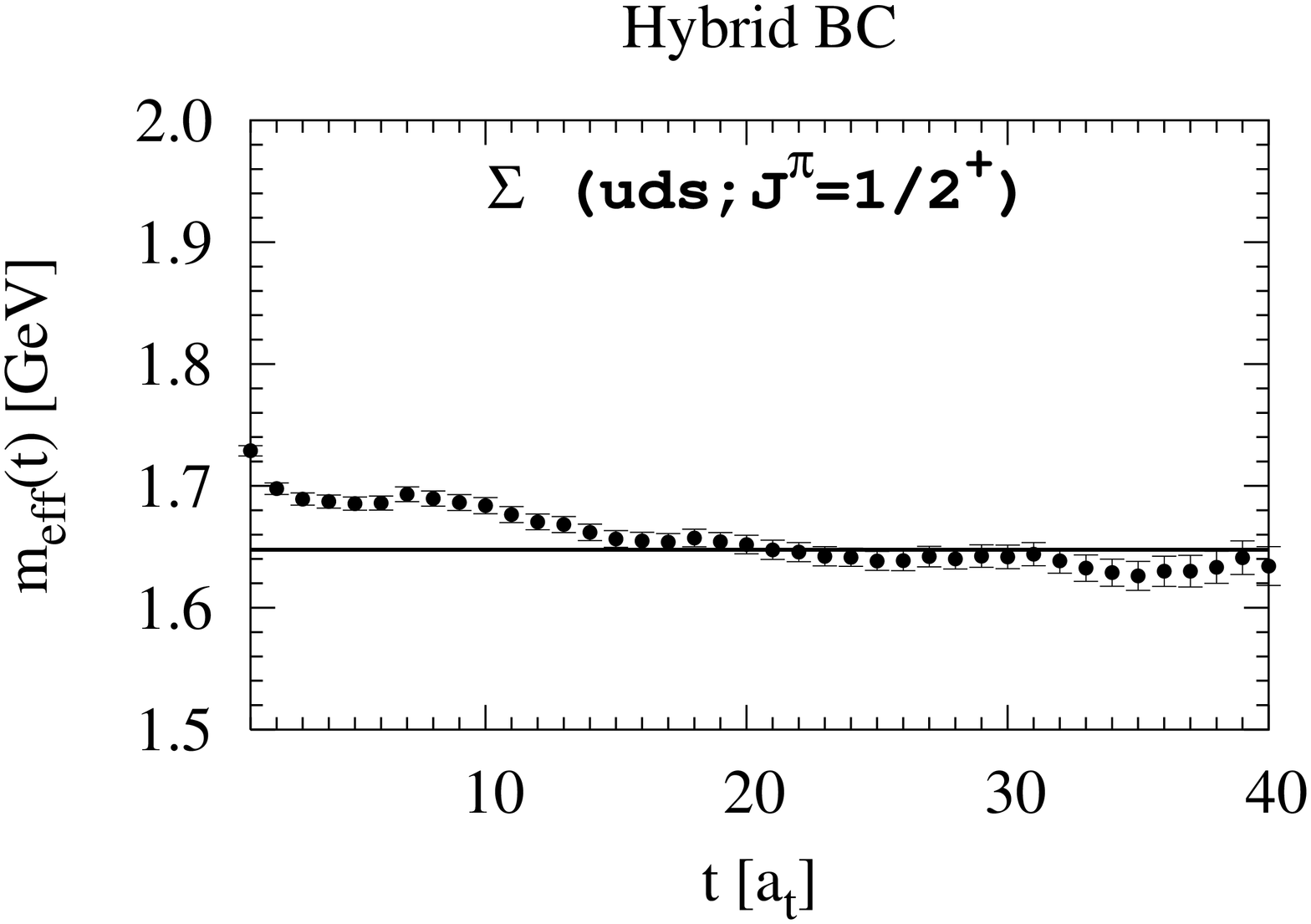,width=0.48\textwidth,angle=-90}
\end{tabular}
\vspace*{-1ex}
\caption{The effective mass plots  of $\Sigma(uds)$ under the standard
BC and HBC adopting \Eq{typical.set}.  Solid lines denote the best-fit
results performed in the region $20\le t \le 30$.}
\label{sigma}
\end{center}
\vspace*{-2ex}
\end{figure}
In \Fig{sigma}, we  show the effective mass plots  of $\Sigma$ for the
standard  BC and  HBC. The  solid  lines denote  the best-fit  results
performed in  the plateau  region as $20  \le t  \le 30$. There  is no
significant difference  between the  two best-fit masses,  which shows
that HBC does not affect the localized resonance states.

\begin{figure}
\vspace*{-3ex}
\centerline{\psfig{file=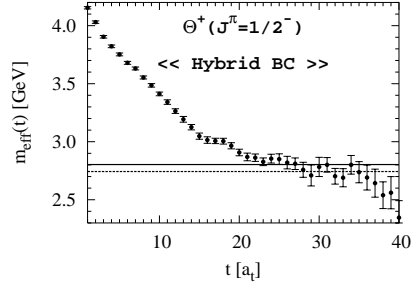,width=0.5\textwidth,angle=-90}}
\vspace*{-1ex}
\caption{The effective  mass plot  for the negative  parity $\Theta^+$
under HBC adopting \Eq{typical.set}. The solid line denotes the result
of  the best-fit  performed in  the plateau  region as  $25 \le  t \le
35$.   The    dotted   line   denotes   the    s-wave   NK   threshold
\Eq{HBC.threshold}. This figure should  be compared with the r.h.s. in
\Fig{fig.effmass}.}
\label{HBC.5Q.effmass}
\vspace*{-2ex}
\end{figure}

Now we present the HBC result of the 5Q effective mass in the negative
parity channel in \Fig{HBC.5Q.effmass}.
The  dotted line  denotes  the  modified NK  threshold  energy due  to
HBC. Note that  the shift of the NK threshold amounts  to about 200 MeV
in  this  case, i.e.,  $L\simeq  2.15$  fm  with \Eq{typical.set}.  We
observe that the  plateau is raised by consistent  amount as the shift
of the threshold, which shows that there is no compact 5Q resonance in
the region as
\begin{equation}
  m_N + m_K
  \le E \le
  \sqrt{m_N^2 + \vec p_{\rm min}^2}
  +
  \sqrt{m_K^2 + \vec p_{\rm min}^2},
  \Hs
  |\vec p_{\rm min}| = \sqrt{3}\pi/L.
\end{equation}
In particular, the plateau observed  in the negative parity channel in
the previous section turns out to be an NK scattering state.

\begin{figure}
\vspace*{-2ex}
\centerline{\psfig{file=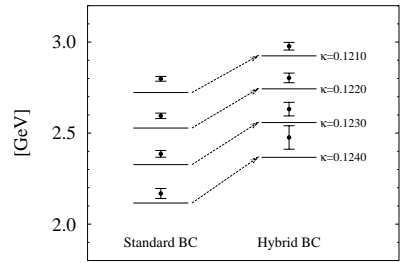,width=0.5\textwidth,angle=-90}}
\vspace*{-1ex}
\caption{Comparison  of the  results of  standard BC  (l.h.s)  and HBC
(r.h.s.) for each $\kappa$.  Dots  denote the best-fit mass obtained in
their  plateau regions. The  solid lines  denote the  corresponding NK
threshold energies.}
\label{hikaku}
\vspace*{-5ex}
\end{figure}

In \Fig{hikaku}, we show the  comparison of the results of standard BC
(l.h.s.) and HBC (r.h.s.) for  each $\kappa$. Dots denote the best-fit
mass obtained in  their plateau region. the solid  lines denote the NK
threshold energies.  We  see that, for all $\kappa$,  the plateaus are
raised  in  a consistent  amount  as the  shift  of  the NK  threshold
energies.
\section{Summary and Discussion}
We  have studied the  penta-quark $\Theta^+$  state with  the quenched
anisotropic lattice QCD to provide high-precision data.
After the chiral extrapolation, we have obtained $m_{\Theta}=2.25$ GeV
in the positive parity channel.
Since  it  is  too  massive,  we  have concluded  that  it  cannot  be
identified with the experimentally observed $\Theta^+(1540)$.
In  contrast,  in  the  negative  parity  channel,  we  have  obtained
$m_{\Theta}=1.75$ GeV,  which is rather close to  the empirical value.
In order to confirm that this state is a compact 5Q resonance, we have
proposed a  new general method  with ``{\em hybrid  boundary condition
(HBC)}''. The HBC analysis has  showed that the observed states in the
negative parity  channel are  NK scattering states  for all  values of
$\kappa$.

We have thus observed no  relevant signals on the compact 5Q resonance
in  both the  parity channels.   To  reveal the  mysterious nature  of
$\Theta^+(1540)$,   more   systematic   investigations  seem   to   be
necessary. In the  future study, it is desirable  to examine the large
volume  effect, the  dynamical quark  effect,  different interpolating
fields including highly non-local  ones, and different quantum numbers
other than $J=1/2$, $I=0$.
%

\end{document}